\documentclass[12pt]{article}

\begin{document}
\bibliographystyle{osa2}
\section*{C-axis Josephson Plasma Resonance Observed in
 Tl$_{2}$Ba$_{2}$CaCu$_{2}$O$_{8}$ Superconducting Thin Films using
 Terahertz Time-Domain Spectroscopy}

\subsection*{V. K. Thorsm$ \o$lle, R. D. Averitt, M. P. Maley,\\
L. N. Bulaevskii, C. Helm, A. J. Taylor} 

\begin{center}
Materials Science and Technology Division, Los Alamos National Lab 
 Los Alamos National Lab, MS K763, Los
Alamos, NM 87545\\
\noindent {\bf{email}}: vthorsmolle@lanl.gov
\end{center}

\begin{abstract}
We have unambiguously observed the c-axis Josephson Plasma Resonance (JPR) in high-T$_{c}$ cuprate Tl$_{2}$Ba$_{2}$CaCu$_{2}$O$_{8}$ superconducting thin films employing terahertz time-domain spectroscopy in transmission as a function of temperature in zero magnetic field. These are the first measurements of the JPR temperature dependence of a high-$T_{c}$ material in transmission. With increasing temperature, the JPR shifts from 705 GHz at 10 K to $\sim$170 GHz at 98 K corresponding to a c-axis penetration depth increasing from 20.1$\pm$0.6 $\mu$m to 84$\pm$9 $\mu$m. The linewidth of the JPR peak increases with temperature, indicative of an increase in the quasiparticle scattering rate. We have probed the onset of the c-axis phase coherence to $\sim$0.95T$_{c}$. The JPR vanishes above T$_{c}$ as expected.
\end{abstract}

310.6860 Thin films and optical properties, 999.9999 Terahertz 
Spectroscopy, 999.9999 Josephson Plasma Resonance, 999.9999
High Temperature Superconductors

\newpage

The c-axis Josephson Plasma Resonance (JPR) in highly anisotropic layered cuprate superconductors originates from the interlayer tunneling of Cooper pairs. Measurements of the JPR in these Josephson-coupled layered systems have received much interest in recent years due to its direct relation to the London penetration depth along the c-axis, $\lambda_{c}$. The JPR, $\omega_{pc}$ = c/$\lambda_{c}\sqrt{\epsilon^{c}_{\infty}}$ = c/$\gamma\lambda_{ab}\sqrt{\epsilon^{c}_{\infty}}$ is thus a fundamental probe of the superconducting state and an excellent tool to study these highly anisotropic systems. Here $\epsilon^{c}_{\infty}$ is the high frequency dielectric constant along the c-axis, and $\gamma$ is the anisotropy parameter. For T$\ll$T$_{c}$, the temperature dependence of $\lambda_{c}$ is related to the symmetry of the order parameter. For T close to T$_{c}$, the appearance of the JPR probes the onset of interlayer phase coherence. Furthermore, the JPR spectral width is a measure of the quasiparticle scattering rate, if intrinsic. In addition, the JPR probes the tunneling mechanism and provides information about the validity of the interlayer tunneling model \cite{vanderMarel98Nature, vanderMarel98ProcSpie}. In a magnetic field, the JPR probes the correlation of pancake vortices along the c-axis and is a tool to study the B-T phase diagram \cite{MartinLev95PRL, Kosugi97PRL}. The advantage of using terahertz time-domain spectroscopy (THz-TDS) to map out the phase diagram is that  this technique allows measurements to be made over a broad frequency range at fixed magnetic field and temperature. The terahertz regime is very important because it overlaps both the JPR and the quasiparticle scattering rates of high-T$_{c}$ superconductors with extreme anisotropy such as the bismuth, thallium, and mercury based high-T$_{c}$ superconductors.

Earlier JPR experiments include microwave cavity experiments on Bi$_{2}$Sr$_{2}$CaCu$_{2}$O$_{8}$ at a fixed frequency on the order of 30-90 GHz utilizing the fact that the microwave absorption resonance peak can be tuned by temperature and magnetic field \cite{Kosugi97PRL, TsuiMatsuda94PRL, Tsui96PRL}. While these measurements demonstrate the existence of the JPR and are useful in probing the system over a narrow range of temperatures and fields, the pancake vortices from the required magnetic field alter the system. Recently, Gaifullin et. al. \cite{Matsuda00PRL} overcame this shortcoming and probed the vortex system of underdoped Bi$_{2}$Sr$_{2}$CaCu$_{2}$O$_{8}$ in a microwave guide by sweeping the frequency continuously from 20 to 150 GHz at constant magnetic field and temperature. However, in order to measure the JPR of less anisotropic high-$T_{c}$ superconductors such as the thallium, and mercury compounds, optical techniques are needed. Tsvetkov et. al. \cite{vanderMarel98ProcSpie} used a grazing angle reflectivity technique and measured the JPR in Tl$_{2}$Ba$_{2}$CuO$_{6}$ and Tl$_{2}$Ba$_{2}$CaCu$_{2}$O$_{8}$ thin films to be $\sim$840 GHz and $\sim$780 GHz, respectively, for T$\ll$T$_{c}$ in zero field. 

In this paper we present experimental data of the c-axis JPR in Tl$_{2}$Ba$_{2}$CaCu$_{2}$O$_{8}$ as a function of temperature in zero field. The JPR is observed at 705$\pm$5 GHz at 10 K and decreases with increasing temperature to $\sim$170$\pm$15 GHz at 98 K corresponding to a c-axis penetration depth increasing from 20.1$\pm$0.6 $\mu$m to 84$\pm$9 $\mu$m. We also observe an increase in the JPR spectral width with increasing temperature, indicative of an increase in the quasiparticle scattering rate. The JPR vanishes above T$_{c}$ as expected. Measuring the JPR in transmission improves the signal/noise ratio allowing measurements of the JPR and its linewidth closer to T$_{c}$ than previous measurements performed in reflection \cite{vanderMarel98ProcSpie}. The onset of the JPR is observed at $\sim$0.95T$_{c}$. The experiments were performed using THz-TDS in transmission with a bandwidth covering approximately 0.2-2.5 THz. The details of the experimental setup are discussed in a previous paper \cite{averittet00JOSAB}. The Tl$_{2}$Ba$_{2}$CaCu$_{2}$O$_{8}$ film (700 nm) was grown on a 10mm$\times$10mm MgO substrate in a two-step process. First an amorphous TBCCO film is deposited by laser ablation. Next it is annealed at high temperature in oxygen to form the 2212 phase. It exhibits a sharp transition (0.2 K width) at a temperature of 103.4 K.

In order to excite the c-axis JPR a component of the electric field (E-field) of the terahertz pulse has to be along the c-axis of the sample. The plasma frequency along the ab-plane is close to the metallic regime $\sim$1.5 eV, while the JPR along the c-axis lies in the terahertz range. Therefore, by tilting the sample at an angle, $\theta$ incident p-polarized THz radiation will be transmitted at the JPR when there is an E-field component parallel to the c-axis and completely reflected when the E-field is parallel to the ab-plane. The anisotropic dielectric function can be written as,
\begin{eqnarray}
\epsilon_{ab}(\omega)=\epsilon^{ab}_{\infty}
\left(
1-\frac{\omega^{2}_{pab}}{\omega^{2}} \right) , \qquad
\epsilon_{c}(\omega)=\epsilon^{c}_{\infty}
\left(
1-\frac{\omega^{2}_{pc}}{\omega^{2}}+\frac{4
{\pi}i\sigma_{c}}{\epsilon^{c}_{\infty}\omega} \right)
\end{eqnarray}
Here $\omega_{pab}$ and $\omega_{pc}$ are the in-plane and out-of-plane plasma frequencies, respectively. Assuming $\omega\ll\omega_{pab}$, and not taking into account dissipation and c-axis dispersion effects \cite{dispersioneffect1, dispersioneffect2}, we find that electromagnetic waves can only propagate in a narrow window above the plasma edge,
\begin{equation}
\omega_{pc}<\omega< \frac{\omega_{pc}}{\sqrt{1- 
\frac{\sin^{2}\theta}{\epsilon^{c}_{\infty}}}}
\approx
\omega_{pc} \left( 1+\frac{\sin^{2}\theta}{2\epsilon^{c}_{\infty}}\right)
\end{equation}
It is a well known fact that the electromagnetic wave in an anisotropic media does not have a pure transverse character, but also has a longitudinal component.

The configuration of our sample with respect to the terahertz beam is shown in Fig. 1. P-polarized THz radiation, incident at an angle of 45$^\circ$ to the surface normal, is transmitted through the sample. In order to minimize the effect of the substrate, two sets of averaged scans are performed at each temperature. The first set on the sample, film plus substrate, and the other set on a bare reference substrate. The Fast Fourier Transform (FFT) of the sample is then divided by the FFT of the reference. This gives the complex transmission coefficient of the film as a function of frequency.

Fig. 2(a)-(d) shows the electric field amplitude of the terahertz pulse in the time domain transmitted through the sample at different temperatures. Fig. 2(a) shows the terahertz pulse transmitted through the sample at 110 K, above the onset of superconductivity. At 90 K, Fig. 2(b), the onset of c-axis coherent tunneling is just perceptible as a slight oscillation following the main pulse. This damped ringing will show as a broad JPR peak in the frequency domain (See Fig. 3), where the damping is due to quasiparticle scattering. Lowering the temperature to 70 K, Fig. 2(c), the terahertz pulse displays an even stronger oscillation due to a reduction in scattering. This results in a narrowing of the JPR peak (See Fig. 3). The terahertz pulse at 10 K, Fig. 2(d), displays a pronounced ringing lasting for at least 20 ps following the main pulse. These fraces reveal how the oscillation frequency increases as the temperature is lowered resulting in a shifting of the JPR to higher frequencies.

Fig. 3 shows the transmission amplitude as a function of frequency for the Tl$_{2}$Ba$_{2}$CaCu$_{2}$O$_{8}$ film for different temperatures. It clearly illustrates the temperature dependence of the JPR, appearing as a sharp peak at $\sim$705 GHz to 680 GHz for low temperatures (10 K to 40 K) and decreasing with temperature to $\sim$170 GHz at 98 K. The JPR vanishes close to 99 K which is about 4 K below T$_{c}$ (See Fig. 3). The spectral width of the JPR peak increases with temperature, indicative of an increase in the quasiparticle scattering rate. The appearance of the JPR as a peak, at the resonance in the transmission spectrum, agrees with the electromagnetic analysis, where propagation of the electromagnetic wave occurs only in a narrow window above the plasma edge. The decrease in the JPR frequency with temperature is consistent with a decreasing superconducting electron density on approaching T$_{c}$ from below.

At 10 K the half-width at half maximum (HWHM) of the JPR is $\sim$45 GHz. It increases with temperature to $\sim$80 GHz at 95 K. This data cannot be used to consistently determine the c-axis conductivity due possible to experimental artifacts or effects of c-axis dispersion neglected in equation (1) for $\epsilon_{c}(\omega)$. The temporal scans of the THz pulses are terminated at 20 ps in order to eliminate the effect of the first Fabry-Perot reflection due to the 1 mm MgO substrate. This might cause an additional broadening of the JPR peaks in Fig. 3 in the temperature range 10 K to 40 K.

Taking, $\epsilon^{c}_{\infty}$=11.3$\pm$0.5 for the high frequency dielectric constant for the electric field along the c-axis \cite{vanderMarel98ProcSpie}, we obtain for the c-axis penetration depth, $\lambda_{c}$(10K)=20.1$\pm$0.6 $\mu$m, and $\lambda_{c}$(98K)=84$\pm$9 $\mu$m. Similar results were obtained on a 600 nm Tl$_{2}$Ba$_{2}$CaCu$_{2}$O$_{8}$ film grown on LaO substrate (T$_{c}$=105 K). The temperature dependence of the JPR frequency is plotted in Fig. 4. It displays the same behavior as the Tl$_{2}$Ba$_{2}$CaCu$_{2}$O$_{8}$ film on MgO substrate just with a higher lying JPR due to a slight difference in the anisotropy. Here we obtain for the c-axis penetration depth, $\lambda_{c}$(10K)=18.2$\pm$0.5 $\mu$m ($\omega_{pc}/2\pi\approx$780$\pm$5 GHz), and $\lambda_{c}$(90K)=31.6$\pm$1.4 $\mu$m ($\omega_{pc}/2\pi\approx$450$\pm$10 GHz).

To summarize, we have applied THz-TDS to measure the c-axis JPR in Tl$_{2}$Ba$_{2}$CaCu$_{2}$O$_{8}$ as a function of temperature. These measurements agree well with previous measurements on Tl$_{2}$Ba$_{2}$CaCu$_{2}$O$_{8}$, using a grazing angle reflectivity technique \cite{vanderMarel98ProcSpie} and we have been able to extend the measurements to temperatures approaching T$_{c}$. We have probed the onset of the c-axis phase coherence to $\sim$0.95T$_{c}$. Future experiments aim to utilize this new capability of measuring the JPR in transmission to probe the B-T vortex phase diagram of anisotropic high-T$_{c}$ superconductors.  

We would like to thank Superconductivity Technologies Inc., Santa Barbara, California, for providing the Tl$_{2}$Ba$_{2}$CaCu$_{2}$O$_{8}$ film grown on MgO. This research was supported by the University of California Campus-Laboratory Collaborations and by the Los Alamos Directed Research and Development Program by the US Department of Energy.

\newpage

\noindent {\bf Fig. 1.\/} Configuration of the thin film plus substrate with respect to the terahertz beam. P-polarized THz radiation is transmitted through the sample.

\vspace{15mm}

\noindent {\bf Fig. 2.\/} Electric field amplitude of terahertz pulse in the time domain transmitted through the Tl$_{2}$Ba$_{2}$CaCu$_{2}$O$_{8}$ film on MgO substrate at different temperatures. (a) 110 K. Terahertz pulse transmitted above T$_{c}$. (b) 90 K. Just below T$_{c}$, a slight oscillation following the main pulse signifies the onset of the JPR. (c) At 70 K, the JPR oscillation is more pronounced. (d) At 10K, the JPR oscillation persists at least 20 ps.   

\vspace{15mm}

\noindent {\bf Fig. 3.\/} Transmission amplitude versus frequency for the Tl$_{2}$Ba$_{2}$CaCu$_{2}$O$_{8}$ film on MgO substrate. The JPR peak broadens and shifts to lower frequencies with increased temperature.
\vspace{15mm}

\noindent {\bf Fig. 4.\/} JPR frequency versus temperature for the Tl$_{2}$Ba$_{2}$CaCu$_{2}$O$_{8}$ film on both MgO (circles) and LaO (triangles) substrate.

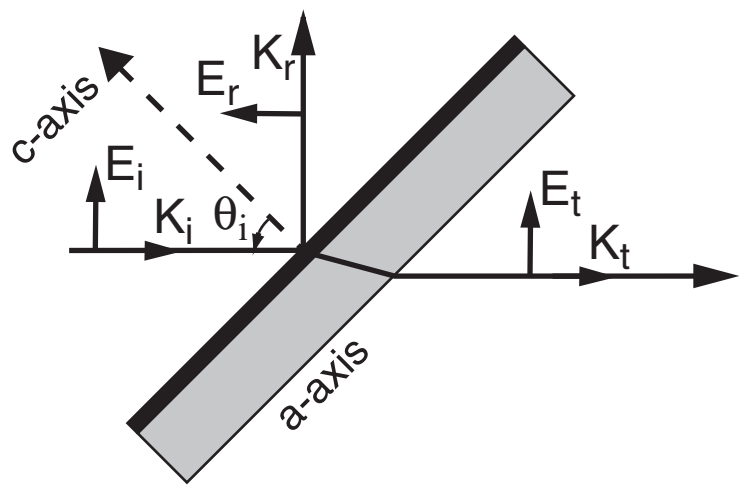
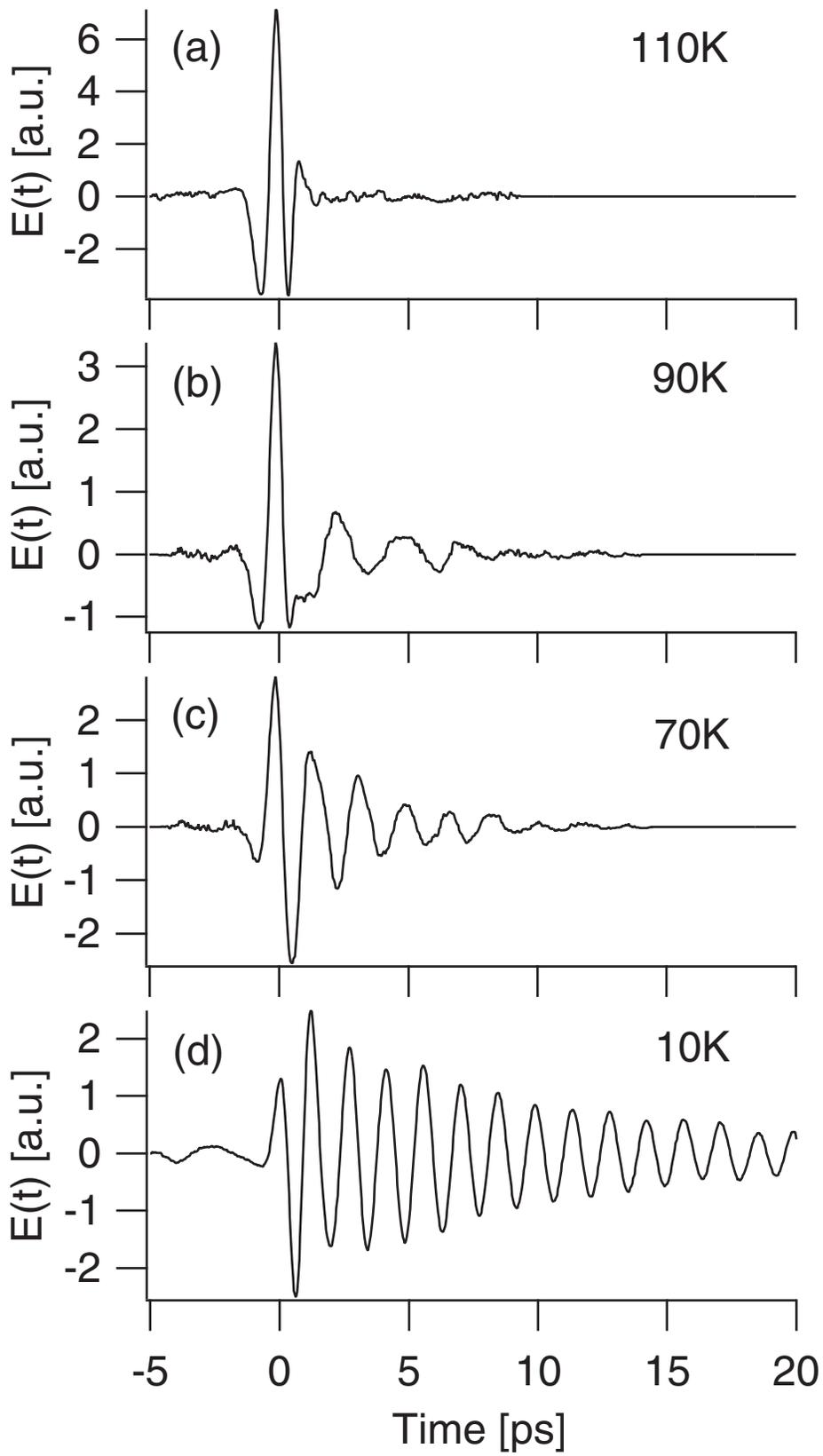
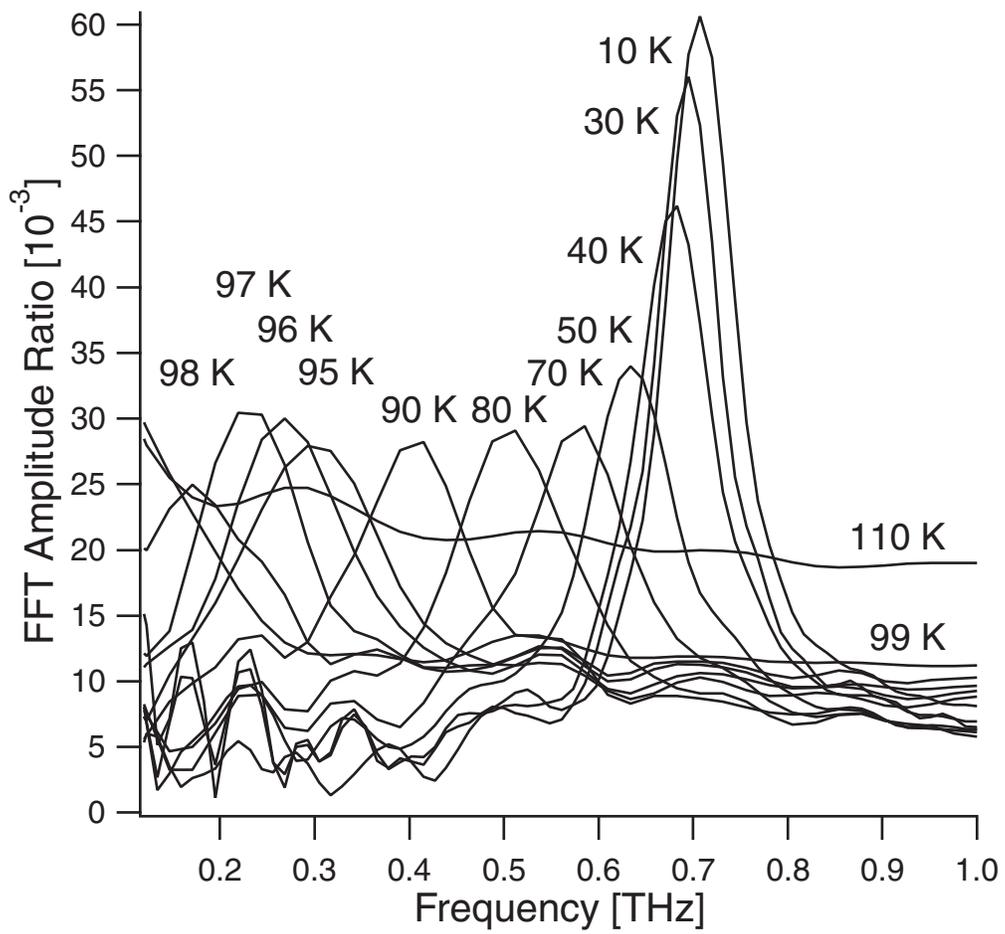
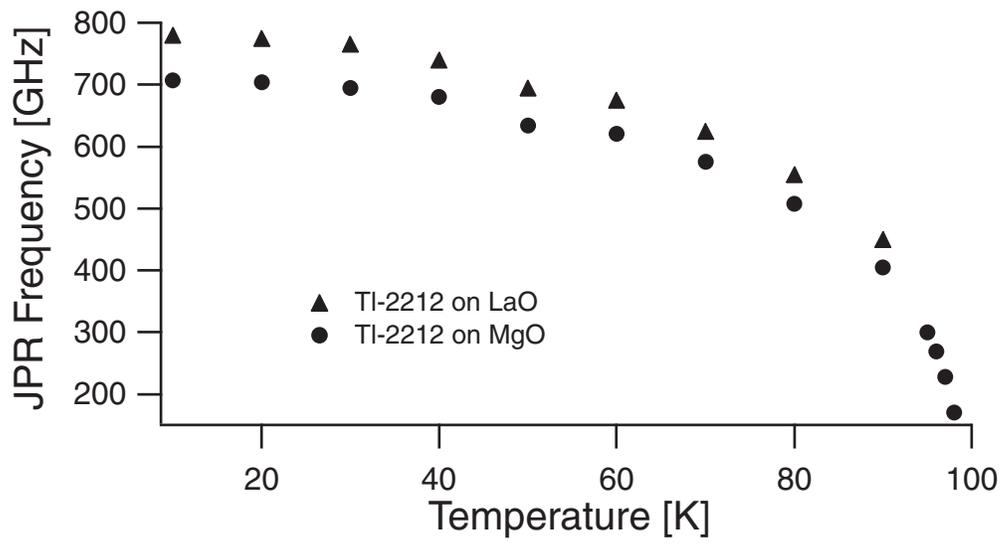
\end{document}